\documentstyle[preprint,aps,psfig]{revtex}
\begin{document}
\draft
\preprint{MKPH-T-96-1}
\title{Unitary ambiguity in the extraction of the 
E2/M1 ratio for the 
$\gamma N\leftrightarrow\Delta$ transition\thanks
{Supported by the Deutsche Forschungsgemeinschaft (SFB 201)}}

\author{P.\ Wilhelm, Th.\ Wilbois, and H.\ Arenh\"ovel}
\address
{Institut f\"ur Kernphysik, Johannes Gutenberg-Universit\"at, 
D-55099 Mainz, Germany}
\maketitle

\begin{abstract}
  The resonant electric quadrupole amplitude in the transition $\gamma
  N\leftrightarrow\Delta(1232)$ is of great interest for the
  understanding of baryon structure.  Various dynamical models have
  been developed to extract it from the corresponding photoproduction
  multipole of pions on nucleons.  It is shown that once such a model
  is specified, a whole class of unitarily equivalent models can be
  constructed, all of them providing exactly the same fit to the
  experimental data. However, they may predict quite different
  resonant amplitudes.  Therefore, the extraction of the E2/M1($\gamma
  N\leftrightarrow\Delta$) ratio (bare or dressed) which is based on a
  dynamical model using a largely phenomenological $\pi N$ interaction
  is not unique.
\end{abstract}



The ratio $R_{EM}$ of the electric quadrupole to the magnetic dipole
amplitude of the $\gamma N \leftrightarrow\Delta$(1232) transition is
an important quantity for our understanding of hadronic structure. It
provides a powerful test for hadron models since it indicates a
deviation from spherical symmetry. For example, in constituent quark
models, it is directly related to the tensor interaction between
quarks.  Consequently, there is considerable experimental effort in
measuring the corresponding $E_{1+}$ and $M_{1+}$ isospin $3/2$
multipole amplitudes for photoproduction of pions on the nucleon
\cite{legs,mainz}. However, all realistic pion photoproduction models
show that both multipoles, in particular $E_{1+}^{3/2}$, contain
nonnegligible nonresonant background contributions. Unfortunately,
their presence complicates the isolation of the resonant parts. 

In the literature, there are basically two different approaches in
order to extract the $\gamma N \leftrightarrow\Delta$ transition
amplitudes.  The first one is the Effective-Lagrangian-Approach (ELA)
adopted by Olsson and Osypowsky \cite{OlO75} and also used later on by
Davidson, Mukhopadhyay and Wittman \cite{DaM91}. In this approach, the
$\pi N$ scattering is not treated dynamically and thus unitarity can
be implemented only phenomenologically using different unitarization
methods (K matrix, Olsson or Noelle prescription) 
which introduces some model dependence. However, in view of the 
phenomenological character of these methods the deeper origin of this 
model dependence remains unclear.

In the second approach, the $\pi N$ interaction is treated dynamically
and thus unitarity is respected automatically.  Various models of this
type have been suggested in the past, e.g., Tanabe and Ohta
\cite{TaO85}, Yang \cite{Yan85}, and Nozawa, Blankleider and Lee
\cite{NoB90}.  However, due to our limited understanding of the
dynamics of the $\pi N$ system, all these models are to a large extent
phenomenological.  Nevertheless, the necessity of such a dynamical
treatment has been stressed again by Bernstein, Nozawa and Moinester
\cite{BeN93}.  Thereby, it is implicitly assumed that the ongoing
improvement of the experimental data base (for both $\pi
N\rightarrow\pi N$ and $\gamma N\rightarrow\pi N$) will finally allow
to favor one of the models and thus will lead to a unique $R_{EM}$.
In this paper we would like to point out an inherent unitary
ambiguity in the latter approach which, to our knowledge, has never
been discussed before.

Qualitatively, one may understand this unitary freedom in the
following way.  First of all, the separation of a resonant $\Delta$
contribution corresponds to the introduction of a $\Delta$ component
into the $\pi N$ scattering state which vanishes in the asymptotic
region.  The explicit form of a wave function depends on the chosen
representation, which can be changed by means of unitary
transformations. As a consequence, the probability of a certain wave
function component is not an observable, because it depends on the
representation.  Classical examples are the deuteron $D$ wave or
isobaric components in nuclei \cite{friar,amado}.  Introducing a
phenomenological $\pi N$ interaction model always implies the choice
of a specific representation.  However, its relation to other
representations and in particular its relationship to hadron models
remains unknown. Thus, it is not clear in principle, how the extracted
resonant multipoles, which are representation dependent, can be
related to the $\gamma N \leftrightarrow \Delta$ transition matrix
elements calculated within, e.g., a nonrelativistic quark model.

\begin{figure}
\centerline{\psfig{file=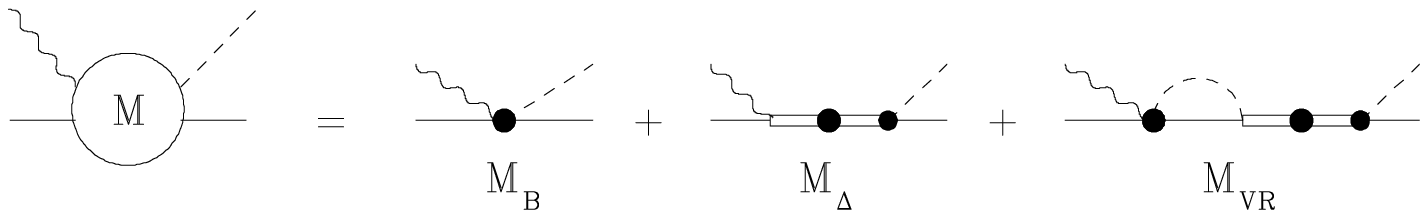,width=10cm,angle=90}}
\caption{The pion photoproduction multipole $M_{1+}^{3/2}$ or $E_{1+}^{3/2}$.}
\label{fig:tgpi}
\end{figure}

We will illustrate our arguments quantitatively by means of a simple 
model \cite{WiW96} whose main features are taken from 
Ref.\ \cite{TaO85}. 
It assumes as Hilbert space  $\Delta
\oplus \pi N \oplus \gamma N$ with corresponding projectors 
$P_\Delta$, $P_{\pi N}$, $P_{\gamma N}$, and a Hamiltonian of the form
\begin{equation}
h = t(m_\Delta ) + v_{\pi\pi}^B + v_\pi + v_\pi^\dagger + 
v_{\pi\gamma}^B +v_\gamma \, ,
\label{eqn:h}
\end{equation}
with the background $\pi N$ interaction $v_{\pi\pi}^B=P_{\pi N} (h-t)
P_{\pi N}$, the $\pi N\Delta$ vertex $v_\pi=P_\Delta h P_{\pi N}$, the
nonresonant $\gamma N\rightarrow \pi N$ driving term
$v_{\pi\gamma}^B=P_{\pi N} h P_{\gamma N}$, and the $\gamma N\Delta$
vertex $v_\gamma =P_\Delta h P_{\gamma N}$.  The kinetic energy $t$ in
the $\Delta$ sector depends on the bare resonance mass $m_\Delta$
which is a model parameter.  The pure hadronic sector ($v_{\pi\pi}^B$,
$v_\pi$, $m_\Delta$) of our model is identical to model B of \cite{TaO85}
and thus yields a good fit of the $\pi N$ scattering phase shift in
the $P_{33}$ channel. The electromagnetic background $v_{\pi\gamma}^B$
is modeled differently in order to guarantee gauge invariance (for
details see \cite{WiW96}).  
For such a dynamical model,
the general structure of one of the
total pion production multipoles $M$ ($M_{1+}^{3/2}$ or
$E_{1+}^{3/2}$) is shown diagrammatically
in Fig.\ \ref{fig:tgpi}.  It consists of three parts,
namely in the notation of \cite{BeN93}, the background $M_B$, the bare
resonant multipole $M_\Delta$, and the vertex renormalization part
$M_{VR}$.  The sum $M_R = M_\Delta + M_{VR}$ is referred to as the
dressed resonant multipole. Formally, one has
\begin{equation}
M=\langle \pi N^{(-)}|v_\gamma+v_{\pi\gamma}^B|\gamma N\rangle_M ,
\end{equation}
with the decomposition $M_\Delta = \langle \pi N^{(-)}|v_\gamma|\gamma
N\rangle_M$ and $M_B+M_{VR} = \langle \pi
N^{(-)}|v_{\pi\gamma}^B|\gamma N\rangle_M$, where $|\pi
N^{(-)}\rangle$ denotes the $\pi N$ scattering state.  The index $M$
on the r.h.s.\ indicates the angular
momentum configuration for the magnetic dipole or the electric
quadrupole absorption of the photon.

Any unitary transformation can be written as $U (\alpha) = e^{i\alpha\chi}$,
with a generator $\chi = \chi^\dagger$ and an arbitrary real number $\alpha$.
Clearly, only generators which are 
nondiagonal with respect to $\Delta \oplus \pi N$ 
have to be considered here.
Keeping in mind that $\chi$ has to be odd under time reversal, a prototype 
is given by
\begin{equation} 
\label{eqn:chi}
\chi = i \,\left[ v_\pi + v_\pi^\dagger ,v_{\pi\pi}^B \right] \, .
\end{equation} 
It obviously mixes resonant and background $\pi N$ interactions and leaves
the $\gamma N$ sector unchanged.
Assuming the background interaction to be of 
separable form, i.e., $v_{\pi \pi}^B = \lambda |b\rangle\langle b|$ with
$\langle b | b\rangle = 1$,
as was actually done in \cite{TaO85,Yan85,NoB90}, $U(\alpha)$
can be evaluated without a perturbative expansion.
Even though the total pion production 
multipole $M$ remains invariant under $U(\alpha)$, its 
decomposition changes according to
\begin{eqnarray}
M_\Delta(\alpha) &=& \langle \pi N^{(-)}|U(-\alpha)P_\Delta U(\alpha) 
(v_\gamma+v_{\pi\gamma}^B)|\gamma N\rangle_M, \\
M_B(\alpha)+M_{VR}(\alpha) &=& \langle \pi N^{(-)}|U(-\alpha)P_{\pi N}
U(\alpha)(v_\gamma+v_{\pi\gamma}^B)|\gamma N\rangle_M.
\end{eqnarray}
Note that by construction $U(\alpha)$ does not modify the initial
state $|\gamma N\rangle$.  Actually, it would be of interest to find
out the representation dependence of both the bare and the dressed
resonant multipole because, as pointed out in \cite{BeN93}, it is
intuitive to compare predictions of nucleon models without and with a
pion cloud to the bare and dressed resonant multipoles, respectively.
In this paper we focus on the bare multipoles. One finds
\begin{equation}
\label{eqn:malpha}
M_\Delta (\alpha) = M_\Delta (0) \left[ 1- 
\frac{1}{2} \left( 1 - r g_M \right) \left( 1 - \cos 2\tilde \alpha \right) +
\frac{1}{2} \left( r + g_M \right) \sin 2\tilde \alpha \right]
\end{equation}
with $g_M = \langle b | v_{\pi\gamma}^B| \gamma N \rangle _M / 
\langle \Delta | v_\gamma | \gamma N \rangle _M $ and 
$r= \langle \pi N ^{(-)} | b\rangle  / \langle \pi N ^{(-)} | \Delta \rangle$,
where $|\Delta\rangle$ is the bare
$\Delta$ state, i.e., $P_\Delta = |\Delta \rangle \langle \Delta |$.
Moreover, we have introduced a
dimensionless parameter ${\tilde \alpha}$ which is proportional to $\alpha$
(for details see \cite{WiW96}). 
Note, that $M_\Delta(\alpha)$ still carries the $P_{33}$ phase shift.
It is easily verified  that, irrespective of the model quantities 
$r$ and $g_M$, 
Eq.\ (\ref{eqn:malpha}) implies that $M_\Delta (\alpha )$
always goes through zero for a certain value of $\alpha$. Consequently, 
the ratio of the bare multipoles $R_{EM}^\Delta  = 
E_{1+,\Delta }^{3/2}/ M_{1+, \Delta }^{3/2}$
as a function of $\alpha $ is in principle {\em unbound}. 

\begin{figure}
\centerline{\psfig{file=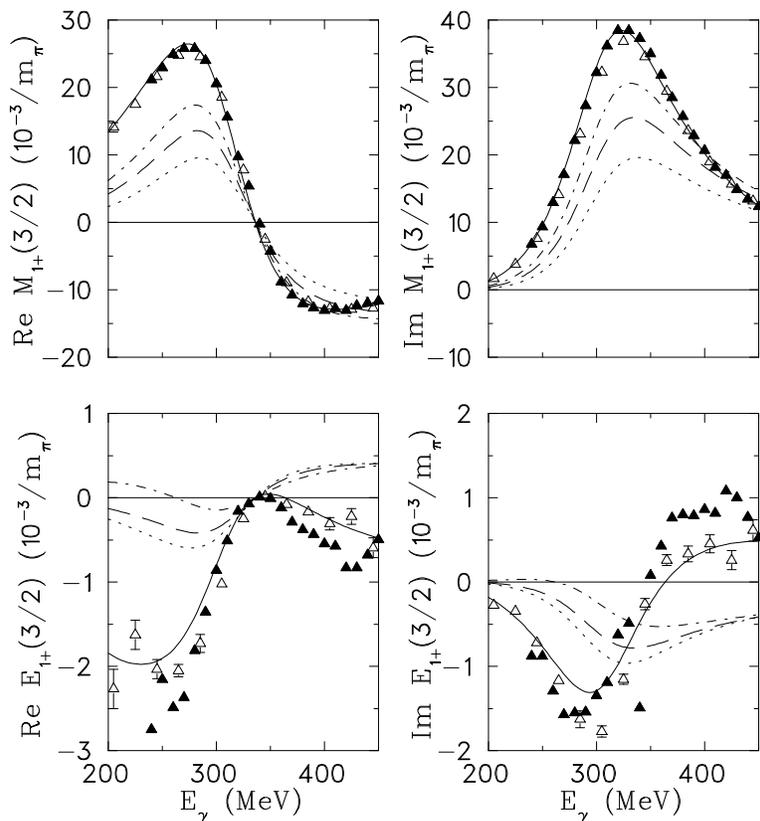,width=10cm}}
\caption{The multipoles $M_{1+}^{3/2}$ and $E_{1+}^{3/2}$ as functions of the 
photon laboratory energy $E_\gamma$.
Dashed, dotted and dash-dotted curves are bare multipoles
corresponding to transformation angles $\tilde \alpha = 0^\circ$,
$10^\circ$ and $-10^\circ$, respectively. The solid curves show the total
multipoles which are representation independent.
The data are taken from \protect\cite{analysis}.}
\label{fig:multipol}
\end{figure}

The representation dependence of both bare multipoles is plotted in
Fig.\ \ref{fig:multipol}. It is already sufficient to consider only
transformations close to the identity.
Even then, the bare amplitudes change substantially,
as can be seen by comparing the dotted ($\tilde \alpha = 10^\circ$)
and dash-dotted curves ($\tilde \alpha = -10^\circ$) with the dashed
one ($\tilde\alpha=0$). For positive $\tilde \alpha$ also
the bare electric multipole exhibits a more pronounced resonance
behavior.  For negative $\tilde \alpha$, the bare
multipoles, in particular $E_{1+,\Delta }^{3/2}$, come
closer to those of Nozawa {\it et al.\ } (see Fig.\ 2 of
\cite{BeN93}).  For completeness we note that the predicted total
multipoles are in satisfactory agreement with experimental results.
Even though we have demonstrated the representation dependence for the
bare multipoles, a similar though weaker dependence occurs also for
the dressed multipoles.

\begin{figure}
\centerline{\psfig{file=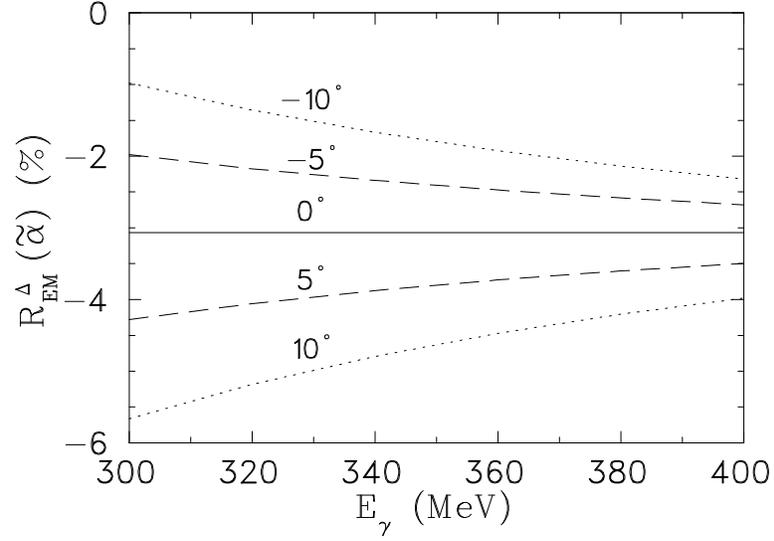,width=10cm,angle=90}}
\caption{The ratio  $R_{EM}^\Delta ({\tilde \alpha} )$  
for ${\tilde \alpha} = 0^\circ ,\pm 5^\circ , \pm 10^\circ$  as 
function of the photon laboratory  energy $E_\gamma$.}
\label{fig:rem}
\end{figure}

The ratio $R_{EM}^\Delta $ is plotted for ${\tilde
  \alpha} = 0^\circ , \pm 5^\circ , \pm 10^\circ$ in Fig.\ 
\ref{fig:rem}. At resonance, it varies strongly between $-1.5$\% and
$-5$\%.  The ratio predicted by our original model is $-3.1\,$\%
\cite{gauge}, which is identical to the result of \cite{NoB90}. The
transformed ratios exhibit a slight energy dependence whereas the
original one is energy independent, which is
just a consequence of the simple ansatz for $v_\gamma$ in \cite{TaO85}
and does not have a deeper physical origin. Moreover, the generated
energy dependence is weak compared to the dependence on
${\tilde\alpha}$.
 
\begin{figure}
\centerline{\psfig{file=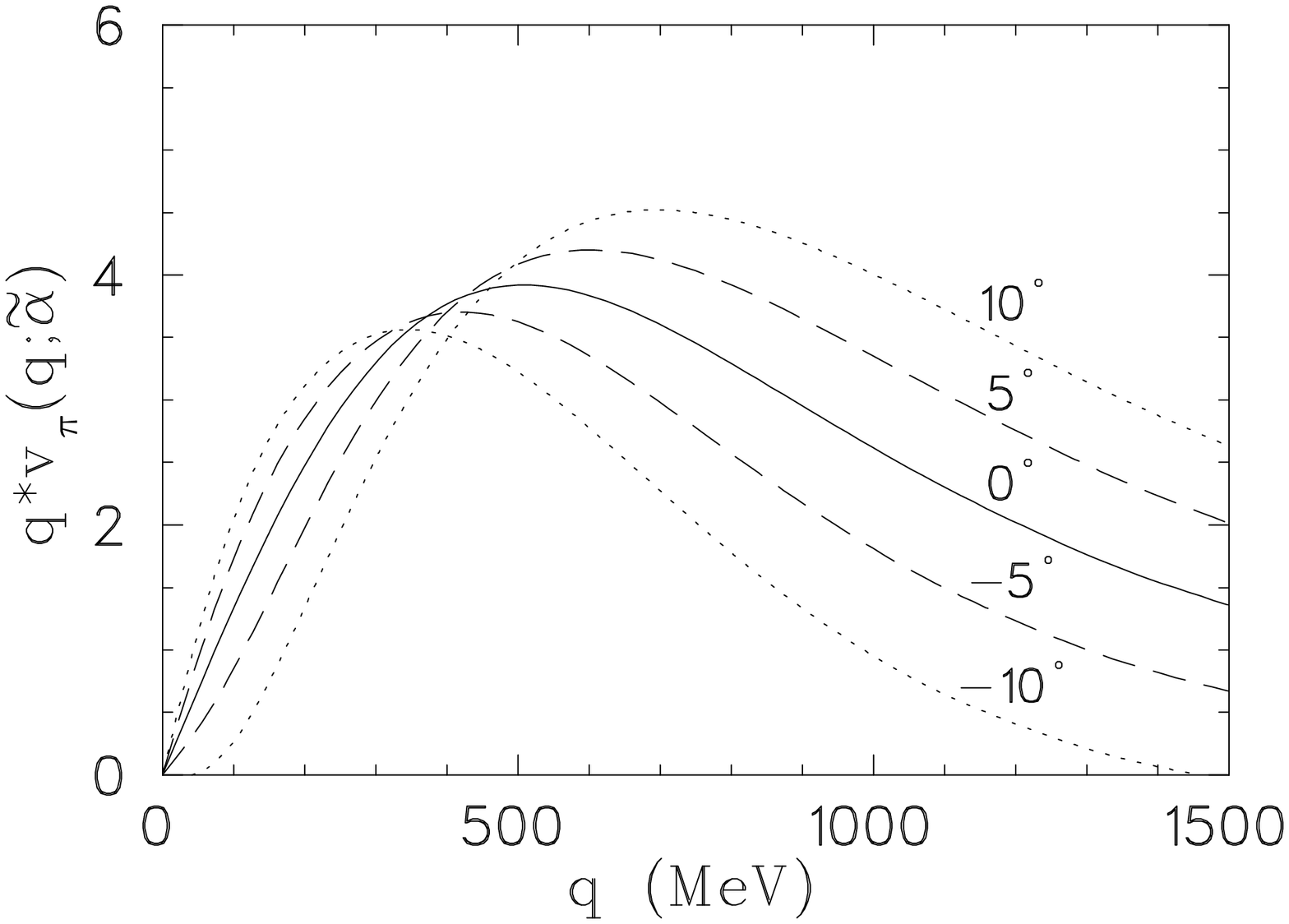,width=10cm,angle=90}}
\caption{The $\pi N\Delta$ form factor $qv_{\pi}(q;\tilde \alpha)$
from Eq.\ (\protect{\ref{eqn:vpi}}) 
for ${\tilde \alpha} = 0^\circ ,\pm 5^\circ , \pm 10^\circ$.}
\label{fig:vpi}
\end{figure}

Now it remains to check whether the transformed Hamiltonian 
$h(\alpha)=U(\alpha)\,h\,U(-\alpha)$ corresponds to 
a ``physically reasonable''
interaction model. Therefore, we write it in the following form
\begin{equation}
\label{eqn:halpha}
h(\alpha) = t(m_\Delta(\alpha)) + v_{\pi\pi}^B(\alpha) 
+ v_\pi(\alpha) + v_\pi^\dagger(\alpha) + 
+ v_{\pi\gamma}^B(\alpha) + v_\gamma(\alpha)\, ,
\end{equation}
where $m_\Delta({ \alpha})  =  \langle \Delta | h({\alpha})| \Delta \rangle$. 
The interaction pieces are defined 
completely analogous to Eq.\ (\ref{eqn:h}), e.g., 
$v_{\pi\pi}^B ({\alpha}) = P_{\pi N}\left( h({\alpha})-t\right) P_{\pi N}$.
The explicit ${\alpha}$-dependence of the various terms is rather
lengthy and will be reported elsewhere \cite{WiW96}.
Since one deals with semiphenomenological interactions here, the only
criterion whether Eq.\ (\ref{eqn:halpha}) is ``physically resonable'' 
will be the
shape of the transformed  form factors. We will demonstrate this 
by considering, for example, the $\pi N\Delta$ form factor. 
Suppressing the isospin structure, it reads
\begin{equation}
\label{eqn:vpi}
\langle\Delta| v_\pi ( \alpha) | \vec q\, \rangle = i \vec S \!\cdot\!
\vec q \, v_{\pi}( q;\tilde \alpha ) \,,
\end{equation}
where $|\vec q\,\rangle $ denotes a plane wave $\pi N$ state with relative
momentum $\vec q$ and $\vec S$ the $N\rightarrow \Delta$ transition
spin operator.  In Fig.\ \ref{fig:vpi}, we have plotted $q\, v_\pi
(q;{\tilde \alpha} )$ for various values of $\tilde \alpha$.
Apparently, none of them can be ruled out.  Here, we just mention that
the modifications of the remaining parts of the transformed
Hamiltonian do not change this conclusion \cite{WiW96}.  Incidentally,
the high momentum components become more and more suppressed when going
from $\tilde \alpha = +10^\circ$ to $-10^\circ$ with a corresponding
increase of $R_{EM}^\Delta$ from $-5$\% to $-1.5$\%.

In summary, we have demonstrated that any extraction of resonant
(bare or dressed) and
nonresonant contributions from the experimental $E_{1+}^{3/2}$ and
$M_{1+}^{3/2}$ multipoles for photoproduction of pions from nucleons
which is based on a dynamical treatment of a  
phenomenological $\pi N$ interaction model suffers
from inherent ambiguities. More precisely, once a phenomenological
model is specified, a whole class of completely equivalent models can
be constructed by means of unitary transformations, all of them
providing {\em exactly} the same fit to the experimental data while
predicting different ratios $R_{EM}^\Delta$. 
Incidentally, also the K-matrix residues in the $\Delta$
region which have been extracted by Davidson and Mukhopadhyay 
\cite{dmk} are not affected by the unitary freedom.

Thus we have to conclude that even with a
perfectly accurate data base one will not be able to discriminate
between any of these models, which actually are merely different
representations. Moreover, those representations which are 
sufficiently close to the original one, are not at all less
``physically reasonable'' because  they cannot be excluded by
arguments based on physical intuition, say, what form factors should
look like.  However, even those models close to the original one
predict significantly different resonant amplitudes. With respect to
our example, none of the different representations, say for
$|{\tilde \alpha}|\!  \leq \!10^\circ$, can be favored, although the
resonant multipoles differ considerably, in particular the ratio
$R_{EM}^\Delta$ varies from $-1.5$\% to $-5$\%.  But one has to keep
in mind, that this variation may be even larger if one considers other
choices for $\chi$ than the one in Eq.\ (\ref{eqn:chi}).  However,
$\chi$ should not contain any quantity which is completely unrelated
to the original Hamiltonian.  

With respect to the model dependence in the ELA, mentioned in the
introduction, it remains to be clarified in the future whether it can
be traced back to unitary transformations relating the different
unitarization procedures.
The arguments presented here may also
affect the separation of resonant and nonresonant amplitudes in other
reactions like, e.g., the particularly interesting $S_{11}$(1535) in
the photoproduction of $\eta$ mesons on nucleons.  Notwithstanding
this unitary ambiguity in the extraction of resonance properties, we
would like to stress the urgent need for more precise data on pion
photoproduction in the $\Delta$ region providing the necessary basis
for an accurate multipole analysis.  However, the challenge is on the
theoretical side because for a clean test of any microscopic hadron
model, the amplitude for the complete process $\gamma N \rightarrow
\pi N$ including background contributions rather than the $\gamma N
\leftrightarrow \Delta$ transition alone, has to be calculated
dynamically within the same model.

\begin{references}
\bibitem{gauge} This value differs considerably from $R_{EM}^\Delta =
  +3.7\,\%$ found in \cite{TaO85}.  The deviation stems from the
  different ansatz for the off-shell behavior of $v_{\pi\gamma}^B$.
  Using the original ansatz of \cite{TaO85}, we can reproduce
  their value.
\end{references}
\end{document}